\documentclass[journal,draftclsnofoot,onecolumn,12pt]{IEEEtran}
\IEEEoverridecommandlockouts
\usepackage{cite}
\usepackage{graphicx}
\usepackage{amsmath}
\usepackage{amssymb}
\usepackage{algorithmicx}
\usepackage{algorithm}
\usepackage{color}
\usepackage{pdfcomment}
\usepackage{bookmark}
\usepackage{enumerate}
\usepackage{bm}

\usepackage{multirow}
\newcommand{\tabincell}[2]{\begin{tabular}{@{}#1@{}}#2\end{tabular}}  
\newcommand{\blc}[1]{\color{black}#1}

\allowdisplaybreaks

\begin{document}
\title{On Channel Reciprocity in Reconfigurable Intelligent Surface Assisted Wireless Network}
\author{
\vspace{1.0cm}
\IEEEauthorblockN{Wankai Tang, Xiangyu Chen, Ming Zheng Chen, Jun Yan Dai,\\ Yu Han, Shi Jin, Qiang Cheng, Geoffrey Ye Li, and Tie Jun Cui}
\thanks{Wankai Tang, Xiangyu Chen, Yu Han, and Shi Jin are with the National Mobile Communications Research Laboratory, Southeast University, Nanjing, China.}
\thanks{Ming Zheng Chen, Qiang Cheng, and Tie Jun Cui are with the State Key Laboratory of Millimeter Waves, Southeast University, Nanjing, China.}
\thanks{Jun Yan Dai is with the State Key Laboratory of Terahertz and Millimeter Waves, City University of Hong Kong, Hong Kong SAR, China.}
\thanks{Geoffrey Ye Li is with the Department of Electrical and Electronic Engineering, Imperial College London, London, UK.}
}
\maketitle
\begin{abstract}
Channel reciprocity greatly facilitates downlink precoding in time-division duplexing (TDD) multiple-input multiple-output (MIMO) communications without the need for channel state information (CSI) feedback. Recently, reconfigurable intelligent surfaces (RISs) emerge as a promising technology to enhance the performance of future wireless networks. However, since the artificial electromagnetic characteristics of RISs {\blc do not strictly follow the normal laws of nature}, it brings up a question: does the channel reciprocity hold in RIS-assisted TDD wireless networks? After briefly reviewing the reciprocity theorem, in this article, we show that there still exists channel reciprocity for RIS-assisted wireless networks satisfying certain conditions. We also experimentally demonstrate the reciprocity at the sub-6 GHz and the millimeter-wave frequency bands by using two fabricated RISs.
Furthermore, we introduce several RIS-assisted approaches to realizing nonreciprocal channels.
Finally, potential opportunities brought by reciprocal/nonreciprocal RISs and future research directions are outlined.
\end{abstract}
\vspace{0.5cm}

%
\IEEEpeerreviewmaketitle

\section{Introduction}\label{Intro}
With the global deployment and commercialization of fifth-generation (5G) mobile communication networks, researchers in both academia and industry around the world are starting to turn their attention to exploring potential enabling technologies for sixth-generation (6G) mobile communication systems. In comparison with the currently deployed 5G networks, 6G is expected to provide much wider geographical coverage, superior energy/spectral efficiency, 10 times higher connection density, lower latency (less than 1 ms), and higher peak data rate (Tera bits per second). With these enhanced performance metrics, it is believed that the fantastic use cases, such as cloud virtual and augmented reality (VAR) based gaming, virtual meeting with holographic projection, and autonomous driving, will be fully supported by 2030 with the aid of 6G networks\cite{6G}.

In order to fulfil the envisioned performance metrics and various applications of 6G, it is necessary to explore new enabling technologies that can provide the performance leaps in 6G. In this context, reconfigurable intelligent surfaces (RISs) have recently been under active investigation as an emerging and promising technology to empower future 6G wireless networks\cite{RISMag}. An RIS is a kind of artificial surface that consists of a large number of sub-wavelength unit cells with tunable electromagnetic responses, including the phase, amplitude, and polarization. For an RIS operating in the microwave band, each of its unit cells is usually made up of the metallic patch, dielectric and tunable component. The reflection coefficient of the unit cell can be adjusted by the control signal applied to its tunable component\cite{RISChinaCom}. By properly tuning the reflection coefficients of the unit cells, RISs are capable of manipulating the wavefront of spatial waves during the wave-matter interaction based on the generalized Snell's law.

RISs are able to perform nearly-passive beamforming on wireless signals {\blc thanks to} their superior and unique capabilities in manipulating spatial electromagnetic waves. In particular, the signals reflected from the unit cells can be co-phased towards the receiver to enhance the quality of received signals\cite{RISTVT}. In addition to the reflection, RISs with specific designs can perform absorption, transmission, and scattering functions on wireless signals, which are specially appealing to applications like wireless power transfer, interference shielding, and secure communications. By using software-defined control strategies and algorithms, RISs have a great potential to proactively improve and customize wireless propagation environments, and flexibly regulate wireless signals according to the desired wireless functions, thus enabling the emerging paradigm known as ``smart radio environments''\cite{RISJSAC}.

The concept of smart radio environments introduces new degrees of freedom to the design and optimization of wireless networks. By deploying RISs on facades and interior walls of buildings, natural wireless channels, which have been regarded as uncontrollable random entities, are expected to be transformed into artificially customized ones\cite{RISTWC}. On the other hand, natural wireless channels in time-division duplexing (TDD) wireless communication networks have the inherent and fundamental property of channel reciprocity. Under a same operating frequency, the access point (AP) can acquire the channel state information (CSI) of the downlink based on the channel estimation of the uplink after some necessary calibrations. This mechanism {\blc minimizes the overhead requirements} for CSI feedback, and thus facilitates the realization of accurate precoding and beamforming in the downlink of TDD wireless networks. However, it is not clear whether the channel reciprocity continues to work in scenarios where RISs are deployed.
This article aims to provide insights on this topic both theoretically and experimentally.

The rest of this article is organized as follows. In Section II, we give a brief review on the reciprocity theorem for traditional wireless channels and unveil the key factors that determine whether the RIS-assisted wireless channel is reciprocal. We then present our experimental results to demonstrate the channel reciprocity of common RIS-assisted wireless networks in Section III. Then, in Section IV, we introduce several approaches to realizing nonreciprocal RISs that break channel reciprocity. We then discuss potential opportunities and future research directions of reciprocal/nonreciprocal RIS-assisted wireless networks in Section V before we conclude the article in Section VI.
\section{On Channel Reciprocity of RIS-assisted Wireless Channels}\label{OnReciprocity}
\begin{figure*}
\centering
\includegraphics[scale = 0.65]{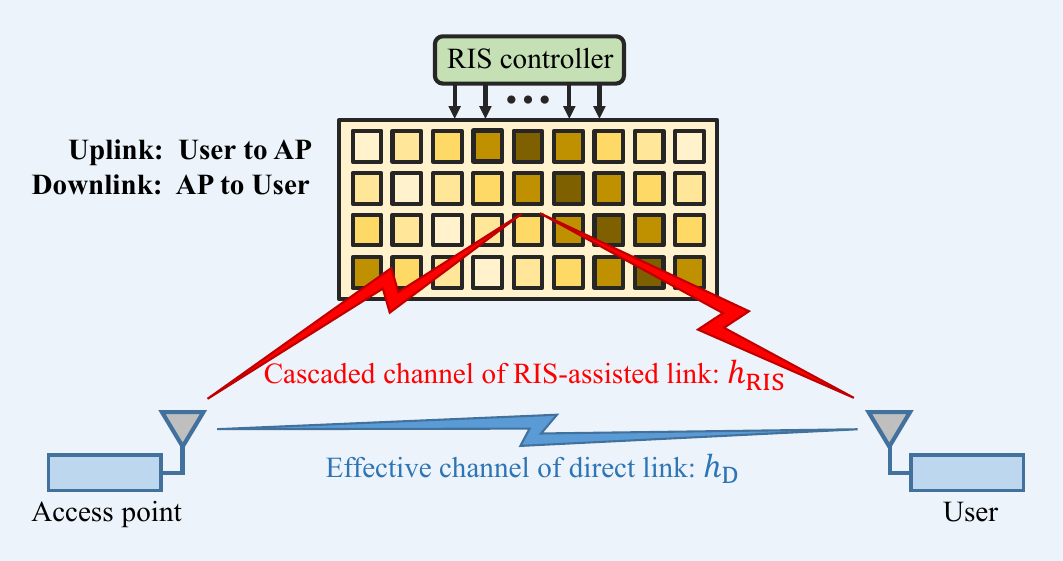}
\vspace{-0.3cm}
\caption{A simple yet typical RIS-assisted TDD wireless communication system.}
\label{System}
\vspace{-0.6cm}
\end{figure*}
Electromagnetic wave is the fundamental basis and physical carrier of modern wireless communications, which makes it possible to realize convenient radio access and communication without the need for wire connections. Electromagnetic wave has many useful physical properties, such as propagating at the speed of light and providing various degrees of freedom. These useful properties endow modern wireless communication systems with advantages, including high mobility, low transmission latency, and superior network capacity.

In this article, we focus on the important property in electromagnetics known as reciprocity, with an emphasis on whether the channel reciprocity holds in RIS-assisted wireless networks. Electromagnetic reciprocity refers to the phenomenon that the electromagnetic field generated at the observation point by a source {\blc remains} the same when the source and observation point swap their positions. This phenomenon comes from the symmetry of Maxwell's equations with respect to time. In particular, as derived from Maxwell's equations for time-invariant and linear media, the most common representation of reciprocity in electromagnetics is the Rayleigh-Carson reciprocity theorem as follows\cite{Origin}
\begin{equation}\label{s1}
\int_{{V_A}} {{{\bf{J}}_A} \cdot {{\bf{E}}_B}d{V_A}}  = \int_{{V_B}} {{{\bf{J}}_B} \cdot {{\bf{E}}_A}d{V_B}},
\end{equation}
where volume $V_i$ contains source $i$ with current density ${\bf{J}}_i$ that creates field ${\bf{E}}_i$. Equation \ref{s1} indicates that the reaction of field ${\bf{E}}_B$ on source A is the same as that of field ${\bf{E}}_A$ on source B, i.e., the interaction between any pair of electromagnetic sources is reciprocal. In particular, Rayleigh-Carson reciprocity theorem unveils that the transmission between a pair of antennas is reciprocal for opposite propagation directions. This property is widely exploited in the field of wireless communications and is known as channel reciprocity.

Without loss of generality, a simple yet typical RIS-assisted TDD wireless communication system is considered in the rest of this article to elaborate on whether the channel reciprocity holds in RIS-assisted TDD wireless networks. As shown in Fig. \ref{System}, the considered system is comprised of a single-antenna access point, a single-antenna user, and an RIS configured by an RIS controller. The wireless channel between the access point and the user can be expressed as
\begin{equation}\label{s3}
h = {h_{\rm{RIS}}} + {h_{\rm{D}}},
\end{equation}
where $h_{\rm{RIS}}$ and $h_{\rm{D}}$ denote the cascaded channel of the RIS-assisted link and the effective channel  of the direct link without the reflection of the RIS, respectively. {\blc $h_{\rm{RIS}}$ is composed of three cascaded parts, namely the channel between AP and the RIS, the reflection coefficient matrix of the RIS, and the channel between the RIS and the user. The reflection coefficient matrix of the RIS is often modeled as a diagonal matrix, whose elements are often phase-adjustable. Therefore, the RIS can flexibly regulate the channel $h_{\rm{RIS}}$ \cite{RISTVT,RISJSAC,RISTWC}.} It is worth noting that the direct link between the access point and the user does not necessarily refer to free-space transmission. It is obvious that the wireless channel of the direct link $h_{\rm{D}}$ is reciprocal because it is a common TDD channel. In other words, the property of the cascaded channel $h_{\rm{RIS}}$ that is provided by the reflection of the RIS determines whether the entire wireless channel of an RIS-assisted wireless network is reciprocal.

In order to better understand whether $h_{\rm{RIS}}$ is reciprocal, let us first revisit the prerequisite of the Rayleigh-Carson reciprocity theorem, which is written as follows\cite{Origin}
\begin{equation}\label{s2}
\begin{array}{*{20}{c}}
{\overline{\overline \varepsilon }  = {{\overline{\overline \varepsilon } }^{\rm T}}}\\
{\overline{\overline \mu }  = {{\overline{\overline \mu } }^{\rm T}}}
\end{array},
\end{equation}
where $\overline{\overline \varepsilon }$ and $\overline{\overline \mu }$ denote the permittivity and permeability tensors, respectively. The tensor identity given {\blc in (\ref{s2})} is used in the process of deriving the Rayleigh-Carson reciprocity theorem from Maxwell's equations. Therefore, this prerequisite unveils that materials and transmission media{\blc ,} for which these conditions expressed in (\ref{s2}) are satisfied{\blc ,} are reciprocal, otherwise they are nonreciprocal. The material tensors of the transmission media and scatterers, such as air, ground, and walls, in the conventional wireless communication systems conform to (\ref{s2}), i.e., the characteristics of the transmission media and scatterers are symmetric for the uplink and the downlink, which is the basic reason and principle for the channel reciprocity in the current TDD wireless networks. As for the RIS-assisted link shown in Fig. \ref{System}, whether the RIS follows conditions in (\ref{s2}) determines whether the cascaded wireless channel $h_{\rm{RIS}}$ is reciprocal.

In general, when the control signals applied to the unit cells remain unchanged, commonly designed and fabricated RISs inherently obey the reciprocity theorem. This is because the materials that constitute RISs, such as metal patches, dielectric layers, and electronic components, often conform to the conditions given in (\ref{s2}).

Nevertheless, there are several special approaches {\blc that break} (\ref{s2}), thereby breaking channel reciprocity in RIS-assisted TDD wireless networks, which will be also discussed subsequently.
\section{Experimental Measurements on Two Typical RISs}\label{Experiment}
In this section, we present the experimental measurements at the sub-6 GHz and millimeter-wave bands to validate whether the channel reciprocity holds in common RIS-assisted TDD wireless communication systems by using two typical fabricated RISs.
\subsection{Two Typical RISs}\label{RISs}
\begin{figure*}
\centering
\includegraphics[scale = 0.35]{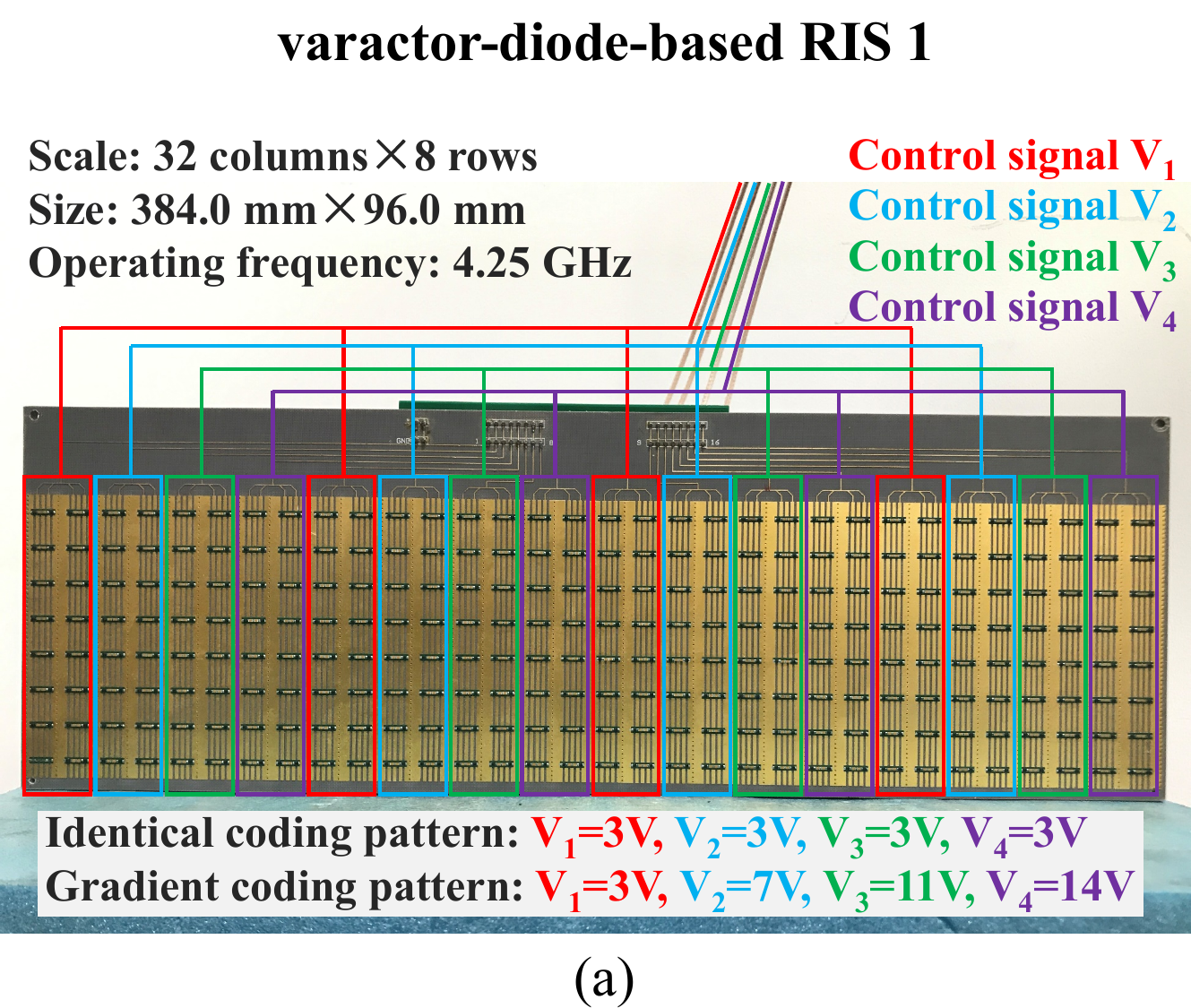}\hspace{0.0cm}
\includegraphics[scale = 0.35]{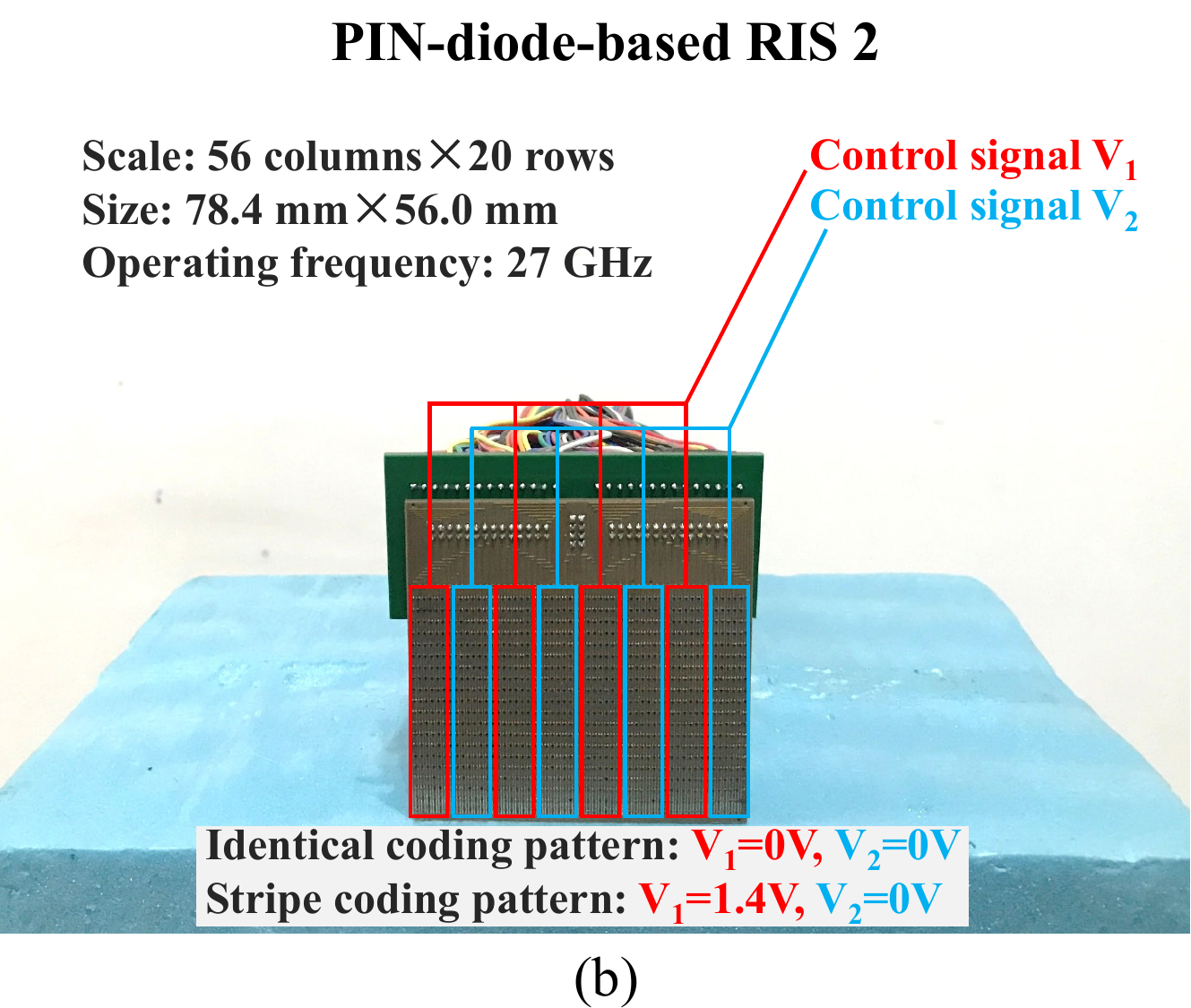}
\vspace{-0.3cm}
\caption{Two typical RISs that are utilized in the measurements. (a) RIS 1 belongs to the varactor-diode-based type. (b) RIS 2 belongs to the PIN-diode-based type.}
\label{FabricatedRISs}
\vspace{-0.6cm}
\end{figure*}
Fig. \ref{FabricatedRISs} illustrates the pictures and parameters of the two typical RISs utilized in the measurements. The word ``typical'' here lies in the fact that the two RISs are commonly designed and fabricated by using standard printed circuit board (PCB) technologies. In other words, no special nonreciprocal materials and designs are {\blc utilized} in these two RISs. In addition, they do cover the most widely used types of RISs, namely varactor-diode-based RISs and positive-intrinsic-negative (PIN)-diode-based RISs. Furthermore, the operating frequencies of these two RISs are in the sub-6 GHz and millimeter-wave frequency bands, respectively. Therefore, the two RISs are sufficiently representative to demonstrate whether the uplink and downlink channels of common RIS-assisted wireless communication systems are reciprocal.

For ease of exposition, the two RISs in Fig. \ref{FabricatedRISs} are referred to as RIS 1 and RIS 2, respectively. As shown in Fig. \ref{FabricatedRISs}(a), RIS 1 belongs to the varactor-diode-based type and works in the sub-6 GHz frequency band. Every two columns of RIS 1 share the same control signal and form a super-column, whose reflection phase can be continuously manipulated as the control signal varies between 0 V and 21 V. Meanwhile, as shown in Fig. \ref{FabricatedRISs}(b), RIS 2 belongs to the PIN-diode-based type and works in the millimeter-wave frequency band. Every seven columns of RIS 2 share the same control signal and form a super-column, whose reflection phase is only 1-bit programmable. More details of these two RISs can be found in \cite{RIS1} and \cite{RIS2}, respectively.

\begin{figure*}
\centering
\includegraphics[scale = 0.85]{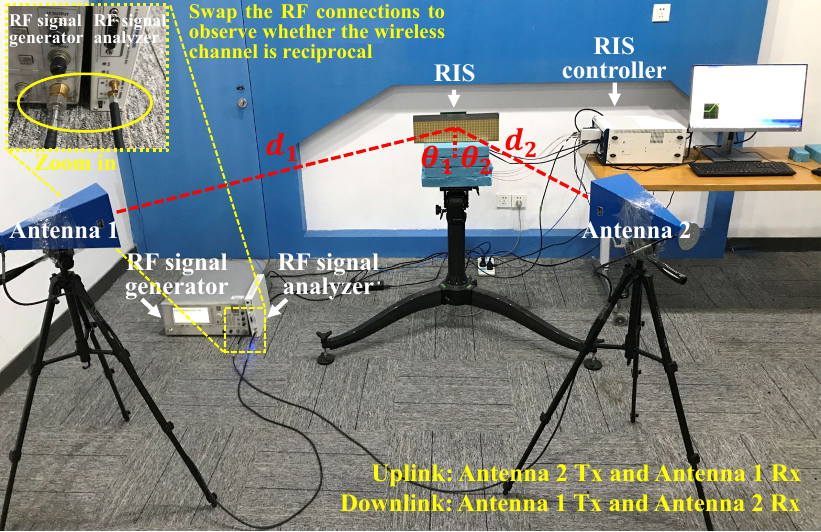}
\vspace{-0.3cm}
\caption{Experiment setup for measuring whether the uplink and downlink channels of an RIS-assisted wireless communication system are reciprocal.}
\label{Testbed}
\vspace{-0.6cm}
\end{figure*}
\subsection{Experiment Setup}\label{Setup}
We have built an experiment setup, as illustrated in Fig. \ref{Testbed}, in order to practically explore whether the reciprocity of RIS-assisted wireless channels holds. In Fig. \ref{Testbed}, $d_1$, $d_2$, $\theta_1$, and $\theta_2$ denote the distance between antenna 1 and the center of the RIS, the distance between antenna 2 and the center of the RIS, the elevation angle from the center of the RIS to antenna 1, and the elevation angle from the center of the RIS to antenna 2, respectively.
{\blc More specifically, $\theta_1$ is the angle between path $d_1$ and the normal of the RIS, and $\theta_2$ is the angle between path $d_2$ and the normal of the RIS.}
The configuration of the RIS is determined by the control signals from the RIS controller. During the measurements, an RF signal generator provides the transmitted signal with power $P_t$ and frequency $f$, and an RF signal analyzer displays the information of the received signal.

As far as the uplink is concerned, antenna 2 is connected to the RF signal generator and antenna 1 is connected to the RF signal analyzer. In this setup, a part of the wireless signal transmitted from antenna 2 is reflected by the RIS and then received by antenna 1, and another part is reflected by the conventional scatterers (e.g., ground and walls) and then received by antenna 1 as well. The wireless channels corresponding to the former and the latter are referred to as ${h_{\rm{RIS}}}$ and ${h_{\rm{D}}}$ as given in equation (\ref{s3}), respectively.

As far as the downlink is concerned, antenna 1 is connected to the RF signal generator and antenna 2 is connected to the RF signal analyzer. Therefore, to switch the experiment setup from the uplink mode to the downlink mode can be simply realized by swapping the RF connections of the RF signal generator and the RF signal analyzer, as shown in Fig.\ref{Testbed}. In consequence, whether the channel reciprocity holds can be proved by observing whether the power (amplitude) and the phase of the received signals in the uplink mode and the downlink mode are consistent with each other.
\subsection{Measurement Results}\label{ExperimentalResults}
\begin{figure*}
\centering
\includegraphics[scale = 0.53]{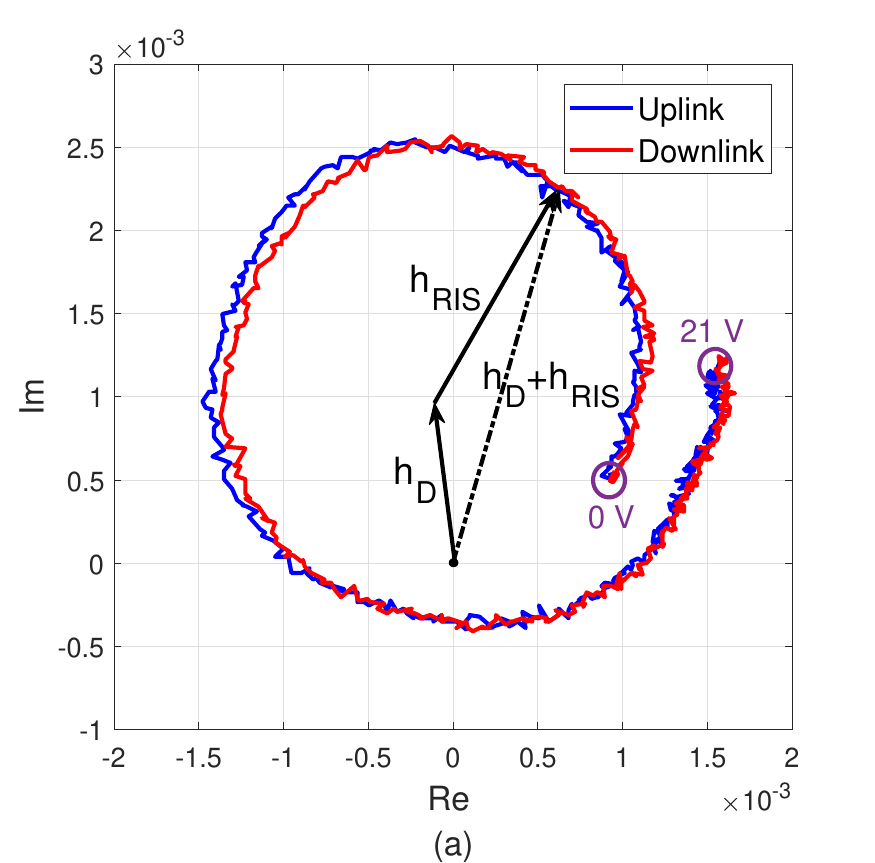}
\includegraphics[scale = 0.53]{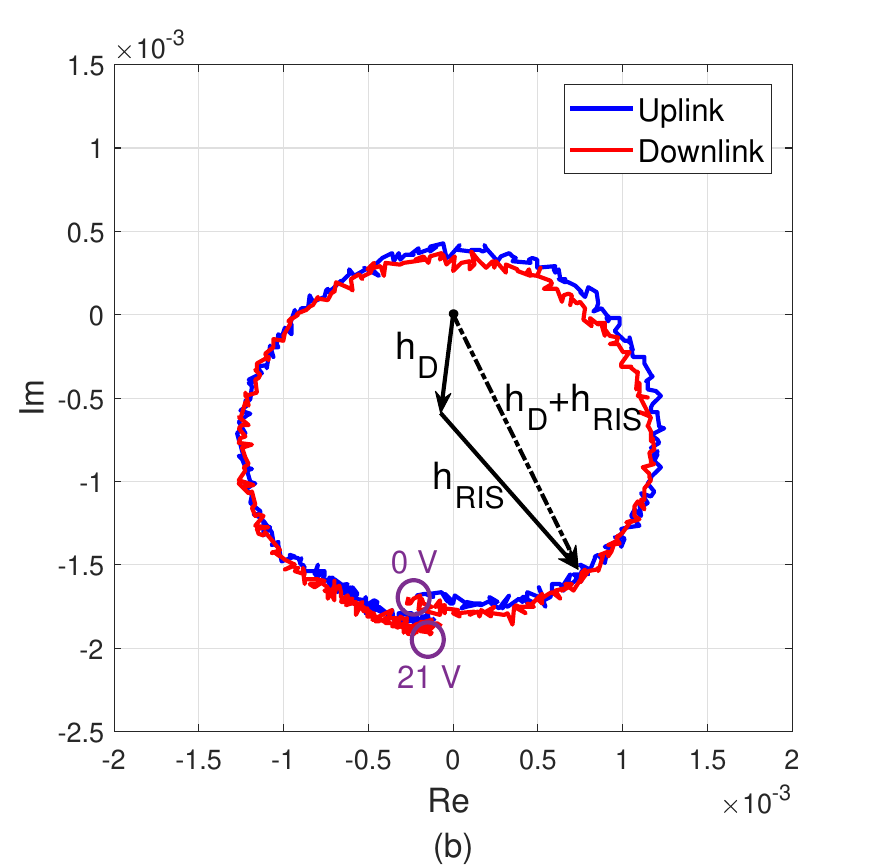}
\vspace{-0.3cm}
\caption{Received signals in the complex plane when a control voltage that varies linearly from 0 V to 21 V is applied to all the unit cells of RIS 1. (a) $d_1=1.5$ m, $\theta_1=30$$^{\circ}$, $d_2=0.5$ m, and $\theta_2=0$$^{\circ}$. (b) $d_1=1.5$ m, $\theta_1=50$$^{\circ}$, $d_2=1$ m, and $\theta_2=30$$^{\circ}$.}
\label{ResultsFigure}
\vspace{-0.6cm}
\end{figure*}
The varactor-diode-based RIS 1 is first employed as the RIS in Fig. \ref{Testbed}. The measurement setup is $P_t=0$ dBm, $f=4.25$ GHz, $d_1=1.5$ m, $\theta_1=30$$^{\circ}$, $d_2=0.5$ m, and $\theta_2=0$$^{\circ}$. By letting the RIS controller apply a control voltage that varies linearly from 0 V to 21 V to all the unit cells of RIS 1 (i.e., let control signals $V_1$, $V_2$, $V_3$, and $V_4$ shown in Fig. \ref{FabricatedRISs}(a) share a same varying voltage), the corresponding received signal in the complex plane is illustrated in Fig. \ref{ResultsFigure}(a). We observe that the received signal moves in the complex plane and forms a trajectory as the control voltage varies from 0 V to 21 V. More specifically, the trajectory of the received signal is actually generated by the sum of $h_{\rm{D}}$ and $h_{\rm{RIS}}$, i.e., $h_{\rm{D}}+h_{\rm{RIS}}$ given in (\ref{s3})\cite{RFocus}. In particular, since the measurement is conducted in a stationary indoor environment, the wireless channel of the direct link $h_{\rm{D}}$ is fixed once the measurement setup is determined. Meanwhile, the wireless channel of the RIS-assisted link $h_{\rm{RIS}}$ is continuously adjusted by the varying control voltage, which leads to the circular trajectory of the received signal shown in Fig. \ref{ResultsFigure}(a). More importantly, from the experiment, the trajectory of the received signal obtained in the downlink mode is in quite good agreement with that obtained in the uplink mode, which indicates that the uplink and downlink wireless channels are reciprocal. For example, as illustrated in Fig. \ref{ResultsFigure}(a), the received signals in the uplink and downlink modes coincide well with each other when the control voltage is 0 V, which demonstrates channel reciprocity.

The same experiment is repeated by considering $d_1=1.5$ m, $\theta_1=50$$^{\circ}$, $d_2=1$ m, and $\theta_2=30$$^{\circ}$. Fig. \ref{ResultsFigure}(b) illustrates the measured received signal in the complex plane. Once again, we observe that the trajectories of the received signals obtained in the downlink mode and in the uplink mode agree well with each other, which proves that the uplink and downlink wireless channels are reciprocal when RIS 1 is employed for RIS-assisted transmission. We should indicate that the trajectories of the received signals in Fig. \ref{ResultsFigure}(a) and in Fig. \ref{ResultsFigure}(b) are different since their corresponding measurement setups, such as $d_1$, $\theta_1$, $d_2$, and $\theta_2$, are different.

\begin{table*}
\centering
\footnotesize
\vspace{-0.2cm}
\caption{Measurement results on the channel reciprocity by using two fabricated RISs.}\label{ResultsTable}
\begin{tabular}{|c|c|c|c|c|}
\hline
\tabincell{c}{\textbf{RIS}} & \textbf{Measurement setup} & \tabincell{c}{\textbf{RIS coding}\\\textbf{pattern}} & \tabincell{c}{\textbf{Received signal power}\\\textbf{(Uplink/Downlink, dBm)}} & \tabincell{c}{\textbf{Received signal phase}\\\textbf{(Uplink/Downlink, degree)}}\\
\hline
\multirow{2}*{RIS 1}  & \multirow{2}*{\tabincell{c}{$P_t=0$ dBm, $f=4.25$ GHz,\\$d_1{=}$1.5 m, $\theta_1{=}$30$^{\circ}$, $d_2{=}$0.5 m, $\theta_2{=}$0$^{\circ}$}} & Identical & -45.5 / -45.5 & 53 / 52 \\
\cline{3-5}
~  & ~ & Gradient & -42.6 / -42.7 & 71 / 70 \\
\hline
\multirow{2}*{RIS 1}  & \multirow{2}*{\tabincell{c}{$P_t=0$ dBm, $f=4.25$ GHz,\\$d_1{=}$1.5 m, $\theta_1{=}$50$^{\circ}$, $d_2{=}$1 m, $\theta_2{=}$30$^{\circ}$}} & Identical & -44.9 / -44.9 & -76 / -77 \\
\cline{3-5}
~  & ~ & Gradient & -53.6 / -53.7 & -139 / -140 \\
\hline
\multirow{2}*{RIS 2}  & \multirow{2}*{\tabincell{c}{$P_t=0$ dBm, $f=27$ GHz,\\$d_1{=}$1 m, $\theta_1{=}$35$^{\circ}$, $d_2{=}$0.5 m, $\theta_2{=}$0$^{\circ}$}} & Identical & -58.0 / -57.8 & NA \\
\cline{3-5}
~  & ~ & Stripe & -49.4 / -49.2 & NA \\
\hline
\multirow{2}*{RIS 2}  & \multirow{2}*{\tabincell{c}{$P_t=0$ dBm, $f=27$ GHz,\\$d_1{=}$1 m, $\theta_1{=}$5$^{\circ}$, $d_2{=}$0.5 m, $\theta_2{=}$0$^{\circ}$}} & Identical & -49.3 / -49.2 & NA \\
\cline{3-5}
~  & ~ & Stripe & -58.2 / -58.2 & NA \\
\hline
\end{tabular}
\vspace{-0.5cm}
\end{table*}

The above-mentioned measurements are carried out under the condition that the control signals of RIS 1 vary dynamically from 0 V to 21 V. In the following, we present the experimental results when the employed RIS is configured by static coding patterns. As illustrated in Fig. \ref{FabricatedRISs}(a), identical coding pattern and gradient coding pattern are applied to RIS 1 in the reciprocity measurements, respectively. Among many possible coding patterns to realize coded RIS 1, identical coding and gradient coding are just two examples that we use to explore the channel reciprocity when RIS 1 is utilized in the experiment. The corresponding measurement results are reported in the first two large rows of Table \ref{ResultsTable}. We observe that the powers (amplitudes) and phases of the received signals in the uplink and downlink are extremely consistent, which unveils that the channel reciprocity still holds. For example, when the measurement setup is $d_1{=}$1.5 m, $\theta_1{=}$30$^{\circ}$, $d_2{=}$0.5 m, $\theta_2{=}$0$^{\circ}$ and gradient coding pattern is applied, the differences between the received powers and the received phases of the uplink and downlink are only 0.1 dB (-42.6$+$42.7$=$0.1) and 1 degree (71$-$70$=$1), respectively. These negligible differences are caused by the slight disturbance in swapping the RF connections during the measurements.

The PIN-diode-based RIS 2 operating in the millimeter-wave frequency band is then employed as the RIS in Fig. \ref{Testbed}. Limited by equipment conditions, we measure and compare the uplink and downlink received signal powers when RIS 2 is configured by identical coding pattern and stripe coding pattern, respectively, as shown in Fig. \ref{FabricatedRISs}(b). The last two large rows of Table \ref{ResultsTable} report the corresponding measurement results. Once again, the received signal powers (amplitudes) in the uplink and downlink agree well with each other, which indicates channel reciprocity when RIS 2 is utilized for RIS-assisted wireless transmission.

{\blc It is worth noting that although the wireless channel of a realistic TDD wireless network is non-stationary, an important prerequisite for channel reciprocity is that the duration of a certain uplink time slot and its corresponding downlink time slot should be shorter than the channel coherence time. Therefore, even though the measurements are conducted in a stationary indoor environment, the experimental results reported above are sufficient to validate that} the channel reciprocity holds in
RIS-assisted TDD wireless communication systems, as long as the employed RISs are commonly designed and fabricated, and conform to the tensor identity given in (\ref{s2}). In addition, Table \ref{ResultsTable} suggests that an appropriate {\blc RIS} coding pattern can significantly increase the received signal power. Moreover, due to the channel reciprocity, the improvements of the received signal powers in the uplink and downlink are always the same.

\section{Breaking Channel Reciprocity}\label{NonreciprocalRISs}
Although channel reciprocity has been often relied upon to design efficient wireless communication protocols, {\blc there are
some potential scenarios where channel reciprocity is not desired, such as wireless power transfer and secure wireless communication, which can be facilitated by breaking the tensor identity in (\ref{s2}).} In this section, we will introduce some approaches to breaking channel reciprocity in RIS-assisted wireless communication systems, including using active nonreciprocal circuits, performing time-varying controls, and employing nonlinearities and structural asymmetries\cite{Microwave}.

\begin{figure*}
\centering
\includegraphics[scale = 0.36]{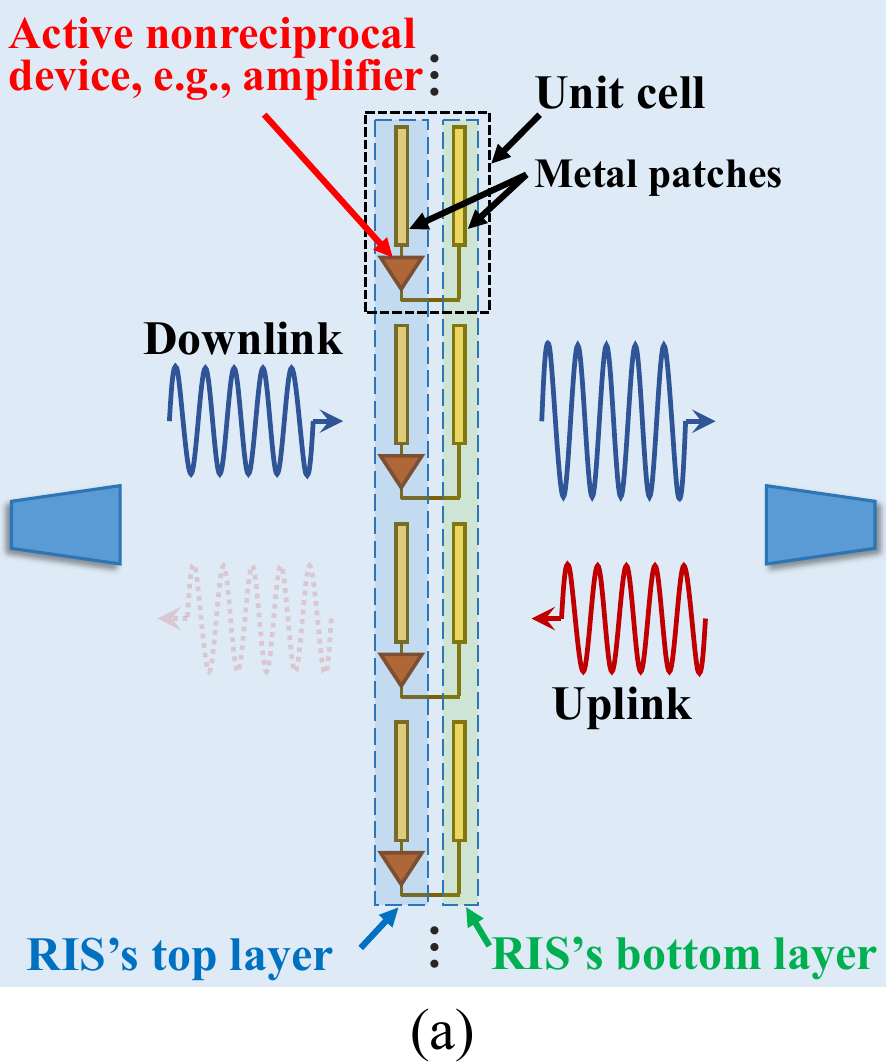}
\includegraphics[scale = 0.36]{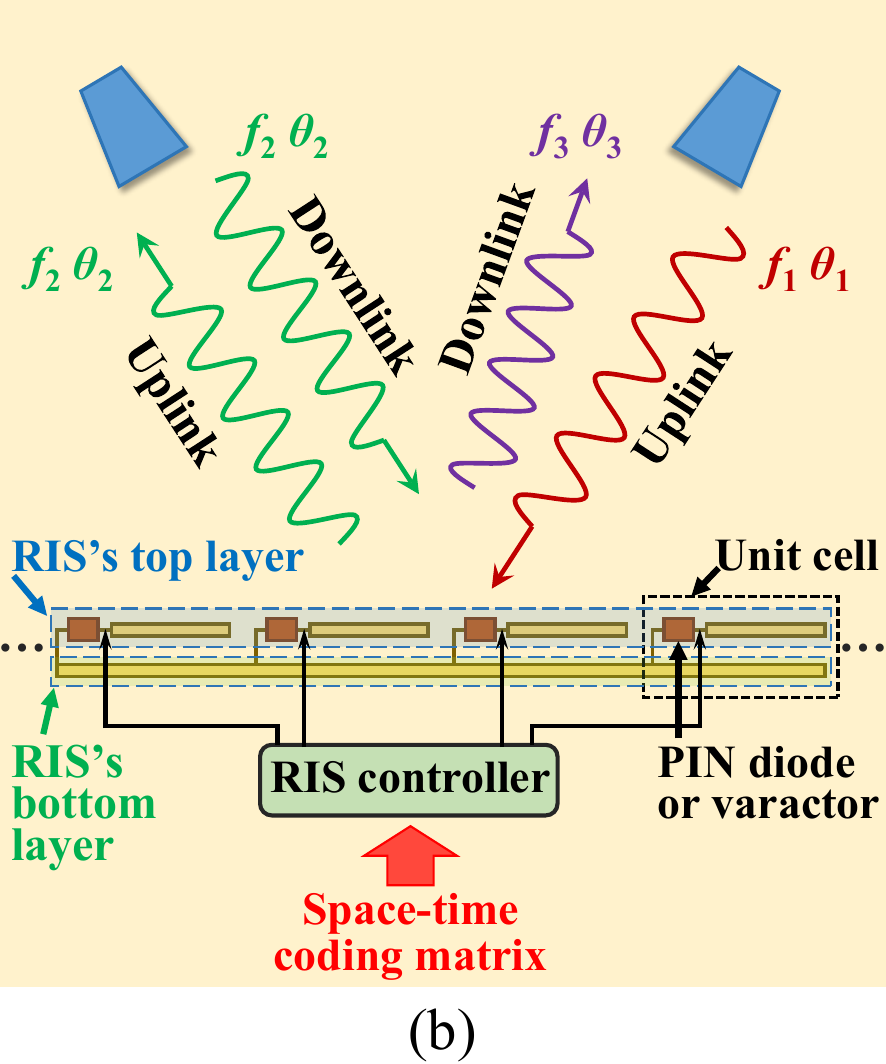}
\includegraphics[scale = 0.36]{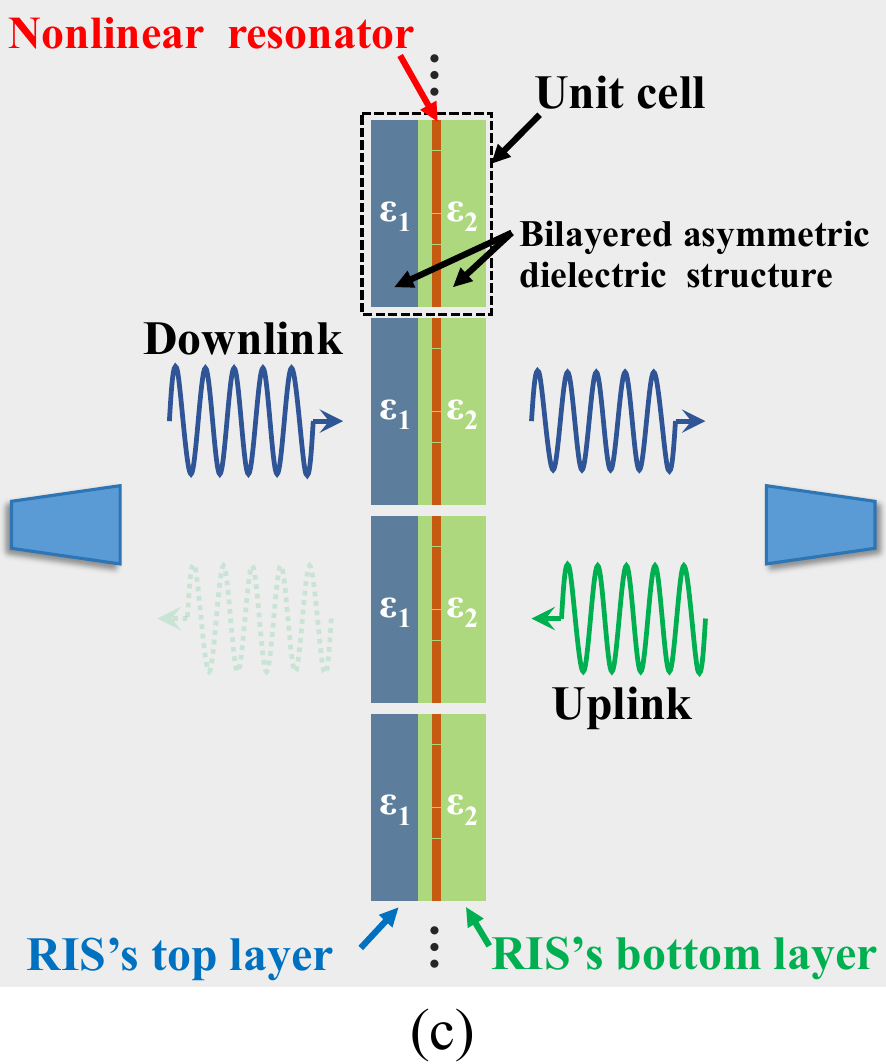}
\vspace{-0.8cm}
\caption{Approaches to breaking channel reciprocity in RIS-assisted TDD wireless networks. (a) Using active nonreciprocal circuits. (b) Performing time-varying controls. (c) Employing nonlinearities and structural asymmetries.}
\label{Approaches}
\vspace{-0.6cm}
\end{figure*}
\subsection{Nonreciprocal RISs Enabled by Active Nonreciprocal Circuits}
By utilizing active nonreciprocal circuits, e.g., microwave amplifiers and isolators, in {\blc the} structure designs of unit cells, nonreciprocal RISs can be achieved with good integrability. For example, an RIS integrated with microwave amplifiers is able to bring about a distinct nonreciprocity between the uplink and downlink wireless channels\cite{MetaAOM}. As depicted in Fig. \ref{Approaches}(a), in the downlink, wireless signals are first captured by the metal patches on RIS's top layer and transformed into circuit signals. After passing through amplifiers and via holes, the circuit signals are then radiated into space again by the metal patches on RIS's bottom layer. In the uplink, however, due to the nonreciprocity of integrated amplifiers, the signals captured by the metal patches on RIS's bottom layer are isolated. In addition, it is worth noting that the varactor-diodes and PIN-diodes, which are often employed in RISs, will not directly cause nonreciprocity, because they are usually only unidirectional for direct current (DC) signals.

\subsection{Nonreciprocal RISs Enabled by Time-varying Controls}
Since the electromagnetic characteristics of RISs are real-time programmable, a straightforward approach to breaking channel reciprocity is to apply different coding patterns to RISs in the uplink and downlink time slots. Moreover, the space-time reconfigurability of RISs further enriches the manifestations of time-varying enabled nonreciprocity. For example, by designing a sophisticated space-time coding matrix applied to an RIS, the wireless signal impinging at an angle $\theta_1$ and frequency $f_1$ can be anomalously reflected at an angle $\theta_2$ and frequency $f_2$ in the uplink, as illustrated in Fig. \ref{Approaches}(b). In the downlink, however, {\blc an} incident wireless signal at an angle $\theta_2$ and frequency $f_2$ is reflected at an angle $\theta_3 \ne \theta_1$ and frequency $f_3 \ne f_1$\cite{MetaAM}. Different space-time coding matrix designs can realize different nonreciprocal features. This example demonstrates that RISs have a great potential to achieve programmable channel {\blc nonreciprocity}.

\subsection{Nonreciprocal RISs Enabled by Nonlinearities and Structural Asymmetries}
Another approach to breaking reciprocity is based on nonlinearities combined with structural asymmetries. For example, the nonreciprocal RIS in \cite{MetaNature} uses a nonlinear resonator in the asymmetric dielectric structure of each unit cell, as shown in Fig. \ref{Approaches}(c). Due to the spatial asymmetry of the two-layer dielectric structure, the amplitudes of the electric fields exciting the nonlinear resonator are different for the uplink and downlink channels, resulting in nonreciprocal transmission properties. As depicted in Fig. \ref{Approaches}(c), the proposed RIS allows wave propagation in the downlink channel, and blocks wave propagation in the uplink channel. Compared with the aforementioned approaches, the resulting RIS needs neither active power supplies nor time-varying control signals to achieve nonreciprocity. Moreover, the uplink transmission coefficient is even related to the power of the incident wireless signals\cite{MetaNature}.

\section{Opportunities and Future Research Directions}\label{Challenges}
We have shown that common RISs are reciprocal because they often inherently conform to the reciprocity theorem. Meanwhile, there are several available approaches to breaking the symmetry of material tensors in order to realize nonreciprocal RISs. In the following, we present some potential opportunities and future research
directions on reciprocal/nonreciprocal RIS-assisted wireless network designs.
\subsection{Signal Amplification Function}
In existing research works {\blc on} RIS-assisted wireless networks, RISs are considered to be nearly-passive, i. e., they have no signal amplification functions.
The amplitude component of each unit cell of an RIS is thus modeled as less than or equal to 1.
However, as mentioned in the previous section, RISs may contain microwave amplifiers, which can provide considerable gains on wireless signals during the wave-matter interaction. For example, a practical transmission gain of about 13 dB is achieved at 5.55 GHz in \cite{MetaAOM}. Therefore, it is necessary to develop {\blc original} signal models for RISs with amplification functions by comprehensively considering the reciprocity, performance, hardware cost, power consumption, and introduced noise. In addition, the RIS reported in \cite{MetaAOM} cannot yet regulate the phase component, and further studies are needed to realize simultaneous phase adjustment and amplitude amplification functions.
\subsection{Switchable Transmission State}
The dramatic increase in wireless devices leads to complex electromagnetic environments, which will cause severe data flow congestion and electromagnetic pollution.
Fortunately, nonreciprocal RISs have demonstrated superior capability in controlling transmission state, which may fundamentally solve these problems. The RIS presented in \cite{MetaAOM} has four switchable transmission states, including bidirectional transmission, forward transmission, backward transmission, and no transmission. Moreover, the building block proposed in \cite{MetaNature} realizes nonreciprocal transmission states. These RIS designs make it possible to directly control data flows from the electromagnetic level while the application cases and analytical models still need further studies.
\subsection{Angle-Dependent Model}
The design methods of RISs are often based on the resonant structures, which inherently makes the reflection/transmission coefficients of the unit cells of RISs angle-dependent\cite{MetaCL}. For example, the phase shift of the unit cells of RIS 1 in Fig. \ref{FabricatedRISs}(a) is highly sensitive to the angle of incidence\cite{RIS1}.
Meanwhile, the measurement results in this article prove that RIS 1 is reciprocal. Therefore, angle-dependent regulation and electromagnetic reciprocity coexist for reciprocal RISs, which unveils that the reflection/transmission coefficient of each unit cell is reciprocal for the exchange of the angle of incidence and the angle of observation. In other words, the reflection/transmission coefficient is not only sensitive to the angle of incidence but also sensitive to the angle of observation. Based on the preliminary angle-dependent phase shifter model proposed in \cite{MetaCL}, {\blc which only considers the dependence on the angle of incidence}, a more tractable and reliable angle-dependent model for reciprocal RISs is urgently needed in order to optimize and predict the performance of RIS-assisted wireless networks.
\subsection{Prototyping and Measurement}
Prototyping and measurement works are the basis of theoretical modeling, link budget analysis, function extension, performance prediction, and configuration optimization of RIS-assisted wireless networks, {\blc and} can help reveal the practical performance limits offered by RISs. At present, studies on the prototyping and measurement of RISs are still in the initial stage and the concept of smart radio environments has not yet been fully demonstrated in practice. It is desired for researchers in academia and industry to jointly explore and promote related works. In addition, currently reported prototyping and measurement works are mostly based on common reciprocal RISs. Whether RISs with signal amplification function and programmable reciprocity can provide performance leaps in RIS-assisted wireless networks needs extensive experiments and measurements to validate\cite{Microwave,MetaAOM,MetaAM,MetaNature}.

\section{Conclusion}\label{Conclusion}
In this article, we have unveiled that the uplink and downlink wireless channels of an RIS-assisted TDD wireless network can be either reciprocal or nonreciprocal, depending on the specific design of the employed RIS. In particular, two typical RISs were utilized in the measurements to validate that the channel reciprocity often holds for commonly designed and fabricated RISs. In addition, we provided an overview of three available approaches to designing nonreciprocal RISs and thus breaking channel reciprocity in RIS-assisted TDD wireless networks. Furthermore, we outlined several future research directions that can facilitate the full potential of reciprocal/nonreciprocal RIS-assisted wireless networks. They include exploring RISs with signal amplification functions, utilizing RISs with switchable transmission states, developing angle-dependent models for RISs, and conducting prototyping and measurement works with both reciprocal and nonreciprocal RISs.

\end{document}